\documentclass[10pt,a4paper]{article}%
\usepackage[top=0.85in,left=1in,footskip=0.75in]{geometry}

\usepackage{changepage}

\usepackage[utf8]{inputenc}

\usepackage{textcomp,marvosym}

\usepackage{fixltx2e}

\usepackage{amsmath,amssymb}

\usepackage{cite}

\usepackage{nameref,hyperref}

\usepackage{microtype}
\DisableLigatures[f]{encoding = *, family = * }

\usepackage{rotating}

\usepackage{soul}

\raggedright
\usepackage[aboveskip=1pt,labelfont=bf,labelsep=period,justification=raggedright,singlelinecheck=off]{caption}

\makeatletter
\renewcommand{\@biblabel}[1]{\quad#1.}
\makeatother

\date{}

\usepackage{lastpage,fancyhdr,graphicx}
\usepackage{epstopdf}
\fancyheadoffset[L]{2.25in}
\fancyfootoffset[L]{2.25in}
\graphicspath{ {./Fig/} }
\def\et{\epsilon_T}

\def\ea{\epsilon_A}

\def\ma{\mu_A}

\usepackage{color}

\definecolor{myred}{rgb}{0.6,0,0}
\definecolor{mygreen}{rgb}{0,0.6,0}
\definecolor{mygray}{rgb}{0.5,0.5,0.5}
\definecolor{mymauve}{rgb}{0.58,0,0.82}

\begin{document}
\vspace*{0.35in}

\begin{flushleft}
{\Large
\textbf\newline{Understanding and coping with extremism in an online collaborative environment}
}
\newline
\\
Csilla Rudas\textsuperscript{1},
Olivér Surányi\textsuperscript{2},
Taha Yasseri\textsuperscript{3,4},
János Török\textsuperscript{1,2,*}
\\
\bigskip
\bf{1} Institute of Physics, Budapest
University of Technology and Economics, Budapest, Hungary
\\
\bf{2} Institute of Physics, Eötvös Loránd University,
Budapest, Hungary
\\
\bf{3} Oxford Internet Institute, University of Oxford, Oxford, UK
\\
\bf{4} Alan Turing Institute, London, UK
\\


\bigskip

%
%



{* torok@phy.bme.hu}

\end{flushleft}
\section*{Abstract}
The Internet has provided us with great opportunities for large scale collaborative public good projects. Wikipedia is a predominant example of such projects where conflicts emerge and get resolved through bottom-up mechanisms leading to the emergence of the largest encyclopedia in human history. Disaccord arises whenever editors with different opinions try to produce an article reflecting a consensual view. The debates are mainly heated by editors with extremist views.  Using a model of common value production, we show that the consensus can only be reached if extremist groups can actively take part in the discussion and if their views are also represented in the common outcome, at least temporarily. We show that banning problematic editors mostly hinders the consensus as it delays discussion and thus the whole consensus building process. To validate the model, relevant quantities are measured both in simulations and Wikipedia which show satisfactory agreement.  We also consider the role of direct communication between editors both in the model and in Wikipedia data (by analysing the Wikipedia {\it talk} pages). While the model suggests that in certain conditions there is an optimal rate of "talking" vs "editing", it correctly predicts that in the current settings of Wikipedia, more activity in talk pages is associated with more controversy.

\section{Introduction}

Large scale collaboration has been a central concept in development of both the Internet and WWW \cite{segal1995short,leiner2009brief,beaudouin1999computer}. With the ever increasing penetration of the information communication technologies across the globe, and the emergence of the user  generated web (sometimes called "Web 2.0"), collaboration of individuals from all around the world to generate public good products is more ubiquitous than ever. A wide range of platforms and protocols facilitate such collaborations between humans and machines at different scales and with different goals \cite{tsvetkova2015understanding}, e.g., Wikipedia, sourceforge, github.

However, in the more and more globalized world of such social systems conflicts may rise due to opinion differences. This is even more important in systems where
common value production is the goal of the service.
In most of the above mentioned systems, due to the bottom-up management in place \cite{Eide2016}, the conventional tools of conflict resolution are inapplicable. 
Hence, it is astonishing that in spite of the magnitude of the opinion
differences in the world even in sensitive issues quality articles are
produced on Wikipedia, comparable to the ones on the expert-written encyclopedias \cite{giles2005internet}. Therefore, the remaining big puzzle about Wikipedia is that ``it only works in practice, in theory, it can never work''.

The approach of complex systems science has become more and more relevant to study collective social behaviour. The availability of large scale data on our personal and societal activities has transformed the methods and scopes of social sciences considerably, leading to the emergence of the new field of computational social science \cite{lazer2009life,watts2013computational}. In this paper, we take such approach and use of agent-based modelling to shed light on some aspects of Wikipedia opinion and content dynamics.

Wikipedia has been studied by various researchers and from different
angles. When it comes to conflicts of Wikipedia, a good amount of
research has been focused on vandalism and how to detect it
\cite{smets2008,potthast2008,wu2010}.
Even though vandalism is very much related to opinion clashes among users, but here we are more interested in conflicts between editors who have faith in the whole project and have no negative incentives. Such cases have been studied empirically by various groups. The bursty nature of editorial wars and the separation between peace and war phases in a dynamical framework, are studied in \cite{dedeo2016} and \cite{yasseri2012dynamics}. More detailed analysis on dyadic interactions between editors and the role of social status are presented in \cite{tsvetkova2016dynamics}. Wikipedia articles have been ranked based on their controversy scores and the controversial topics have been analysed in \cite{vuong2008,yasseri2014most,kittur2009}. And finally tools for visualizing and detecting Wikipedia conflicts are developed \cite{borra2015societal}. However, most of the empirical work on Wikipedia conflicts fail to explain the mechanistic scenarios driving the emergence and resolution of conflicts among editors. 

One of the under-researched aspects of Wikipedia edit wars is the role of the "talk pages". Talk pages are forums in which editors can discuss their opinions on the content of the article and try to reach a consensus before overriding each other's edits directly on the article  \cite{wiki-talk}. Even though it has been argued that the presence of such facilities would hinder the emergence of edit wars \cite{viegas2007}, there is little theoretical work to explain this observation. In other related work, the content of the talk pages is analysed using natural language processing tools to explain their functionality, however again, not much of mechanistic modeling is provided \cite{yasseri2012c,laniado2011a,iosub2014emotions}.

Modelling opinion dynamics in an agent-based framework has an
extensive literature (for a review see
\cite{castellano2009statistical}). A successful class of such models
are known as ``bounded confidence'' models, which allow agents to
accept opinion alterations within a tolerance threshold
\cite{deffuant2000}. In more recent work, bounded confidence models
are generalized to account for emotion dynamics parallel to opinion
dynamics \cite{sobkowicz2015}. Different directions of generalization
of such models have been taken to explain the user dynamics and
opinion dynamics on Wikipedia \cite{torok2013,ciampaglia2011bounded}.
We make use of one of this generalizations which accounts for the
common product (the article) among the agents as well as the indirect
interactions between agents through this common product
\cite{iniguez2014}.

Previous work has shown that in
many cases a consensus can be reached even if the original pool
of opinions was very mixed. It is clear that the major problem in
building a consensus is the presence of extremists who have opinions
very different from the majority. We devote this paper to the study of
this question, that in what extent are extremists impedimental in
consensus building and what measures can be applied to decrease the
chance of a frozen conflict.We also
implement the process of banning extremists editors,  a well known
procedure in Wikipedia and show the effect of it on the evolution of
the conflict. The results of the model are compared with empirical
data generated based on from Wikipedia activity logs.

\section{Methods}
In this paper we combine the editorial activity data collected from 
Wikipedia with a generalized version of the computational model of opinion dynamics that we have developed earlier \cite{torok2013,iniguez2014}. 

\subsection{Data}
The data collection is mainly carried out using Wikimedia Tool Labs\footnote{\url{https://tools.wmflabs.org}}, which provide live access 
to Wikipedia database containing logs of all the editorial activities. For more details on data collection see \cite{yasseri2013}.

In obtaining statistics of the editors' activities, we explicitly excluded Wikibots (semi-automated computer codes that carry out large scale simple tasks, e.g., correcting typos or creating inter-language links). We collected data from different language editions. These editions represent a large range of language editions in size and number of articles, as well as large variety in their local rules and conventions.
However, as reported below, most of the observed statistical features are language independent. 

We counted the number of edits to article pages and Wikipedia talk pages and compared this ratio to its equivalent in the model. To be able to study the consensus reaching process, we consider reverts: edits that undo a previous edit. For more details see \cite{sumi2011a,sumi2011b}.

\subsection{Computational model}

We use the model introduced in \cite{torok2013} and further developed in 
\cite{iniguez2014}. We consider the case with fixed agent pool: $N$
agents try to edit and eventually agree on a descriptive
article about a subject. Each agent has an opinion about it which is
represented by a time-varying scalar variable in the range $x_i(t)\in[0,1]$.
The article can also be biased towards an opinion value at any time represented on the same scalar interval $A(t)\in[0,1]$. At each
timestep a randomly chosen agent with probability $r$ tries to
communicate with another randomly chosen agent or otherwise (with probability
$1{-}r$) tries to edit the article. The model for agent-agent interaction is known as the
Deffuant model for opinion dynamics in mixed populations \cite{deffuant2000}. 

Two agents can only communicate if their opinion differs less than
$\et$, the tolerance parameter of the agent-agent interaction in which case they modify their view
on the subject and both adapt a joint opinion half way between their
original ones:
\begin{equation}\label{Eq:agentagent}
(x_i,x_j)\to\begin{cases}
([x_i+x_j]/2,[x_i+x_j]/2) & \mathrm{if~}|x_i-x_j|<\et\cr
(x_i,x_j)  & \mathrm{otherwise}
\end{cases}
\end{equation}
In the original opinion models \cite{deffuant2000} the only interaction is the agent-agent communication which drives the system
into stable configuration characterized by opinion groups which do not interact with each other, and the average number of which is determined by $\et$.

Agents have different tolerance ($\ea$) towards the opinion reflected in the article. If they
try an editing action and find that the position of the article
differs less than $\ea$ from their own opinion, they do not change it, instead adapt their
opinion towards it by an amount proportional to a convergence
parameter $\ma$ and the opinion difference. In the opposite case when
the article is intolerable for the agent it will modify it accordingly:

\begin{equation}
(x_i,A) \to \begin{cases}
(x_i+[A-x_i]\ma,A) & \mathrm{if~}|x_i-A|<\ea\cr
(x_i,A+[x_i-A]\ma)  & \mathrm{otherwise}
\end{cases}
\end{equation}

\begin{figure}
\begin{center}
\includegraphics[width=0.85\linewidth]{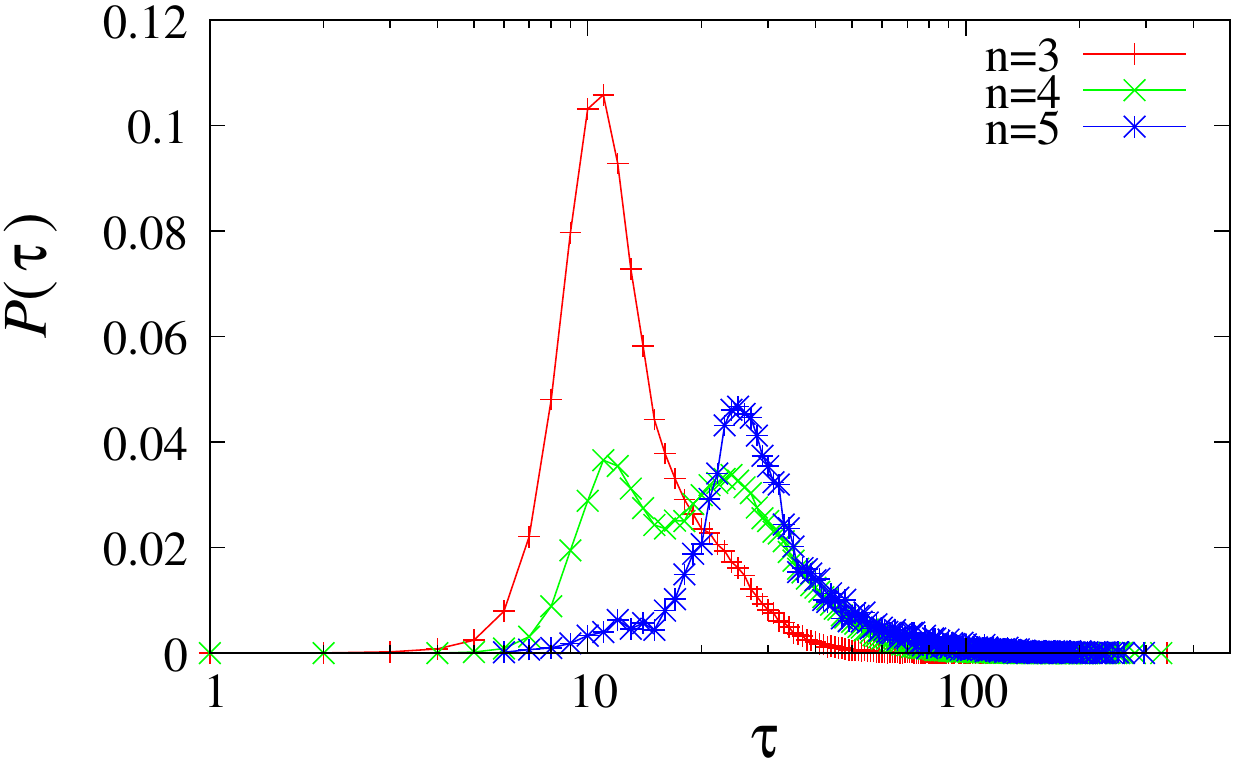}
\caption{\label{Fig:relaxdist} 
Distribution of relaxation time for different number of initial
opinion groups for $\ma=0.65$ and $\ea=0.25$.
}
\end{center}
\end{figure}

The simulation procedure is thus the following: The system is prepared
first by running the agent-agent interaction Eq.~(\ref{Eq:agentagent}) to create the opinion groups, 
then each Monte-Carlo step is composed of $N$ actions in
which a randomly chosen  agent $i$ either talks to another agent or interacts with the 
article as described above.
The relaxation time in general is defined as 
the average number of Monte-Carlo steps needed to reach a
consensus where all agents are satisfied with the article. To cope with the enormous statistical fluctuations which prevent the calculation of a sensible mean, we measured $\tau$ by the position of the maximum of the
relaxation time density function, which is equivalent to the most probable relaxation time.

If the simulations are started with random initial agent opinion
distribution than the number of opinion groups will vary due the intrinsic randomness in the model.
The relaxation
time measured for a specific set of parameters is the average which
includes qualitatively different scenarios of different number of
opinion groups. This phenomenon
illustrated in Fig.~\ref{Fig:relaxdist} where the relaxation time
distribution is shown for simulations with exactly the same parameters but with different random seeds creating different initial opinions $x_i(t{=}0)$.
The distribution of the relaxation time is plotted separately for different number
of initial opinion groups ($n$). Interestingly if initially there are
four opinion groups ($n{=}4$) then the system behaves similar to either a 3 or a 5 group setup.

To avoid the above artifacts emerging from this sensitivity, 
we fix the initial conditions throughout the analysis as the following:

\begin{itemize}
\item {\bf 2 groups:} 2 groups at opinions $0.0$ and $1.0$.
\item {\bf 3 groups:} 1 mainstream group at $0.5$ and 2 extremists
at $0.1$ and $0.9$.
\item {\bf 4 groups:} 2 mainstream groups at $0.25$ and $0.75$,
and 2 extremist groups at $0.0$ and $1.0$.
\end{itemize}
A new parameter $RoE$ is introduced as the ratio of the agents in the extremist groups. Naturally it is relevant only for 3 and 4 groups.

It was shown in \cite{torok2013} that the above defined model has three different
modes of convergence which can be identified by regions in the phase
diagram of $(\ma,\ea)$. We reiterate here the main findings: Regime I
was observed for low values of $\ma$, $\ea$ and was characterized by
astronomical relaxation time (prohibits its study for reasonable system size $N>100$) and an ever lasting stable conflict in which a large mainstream group fights an endless war against two small extremist groups.
 
Therefore we omit the study of this regime here. Regime II was
characterized by an oscillatory behavior of the opinion of the article
and the convergence was reasonably fast. Regime III showed the behaviour most similar to Wikipedia, with very volatile article behavior, and in
parallel, extremists gradually converted to the mainstream opinion.
In this study we will focus on Regime II and III and use the following
parameter values for the simulations:
\begin{itemize}
\item {\bf regime II:} $\ma=0.45$, $\ea=0.075$
\item {\bf regime III:} $\ma=0.7$, $\ea=0.15$
\end{itemize}
Regime II is characterized by a very controversial topic ($\ea$ small) and a moderately volatile article ($\ma$ intermediate),
while Regime III is less controversial ($\ea$ is larger) and more
volatile ($\ma$ is large). We also note here that if $r$ is not too
small than the opinion groups stay compact during the simulations
\cite{torok2013}.


\section{Results and Discussion}
\subsection{Role of the extremists}

\begin{figure}
\begin{center}
\includegraphics[width=0.85\linewidth]{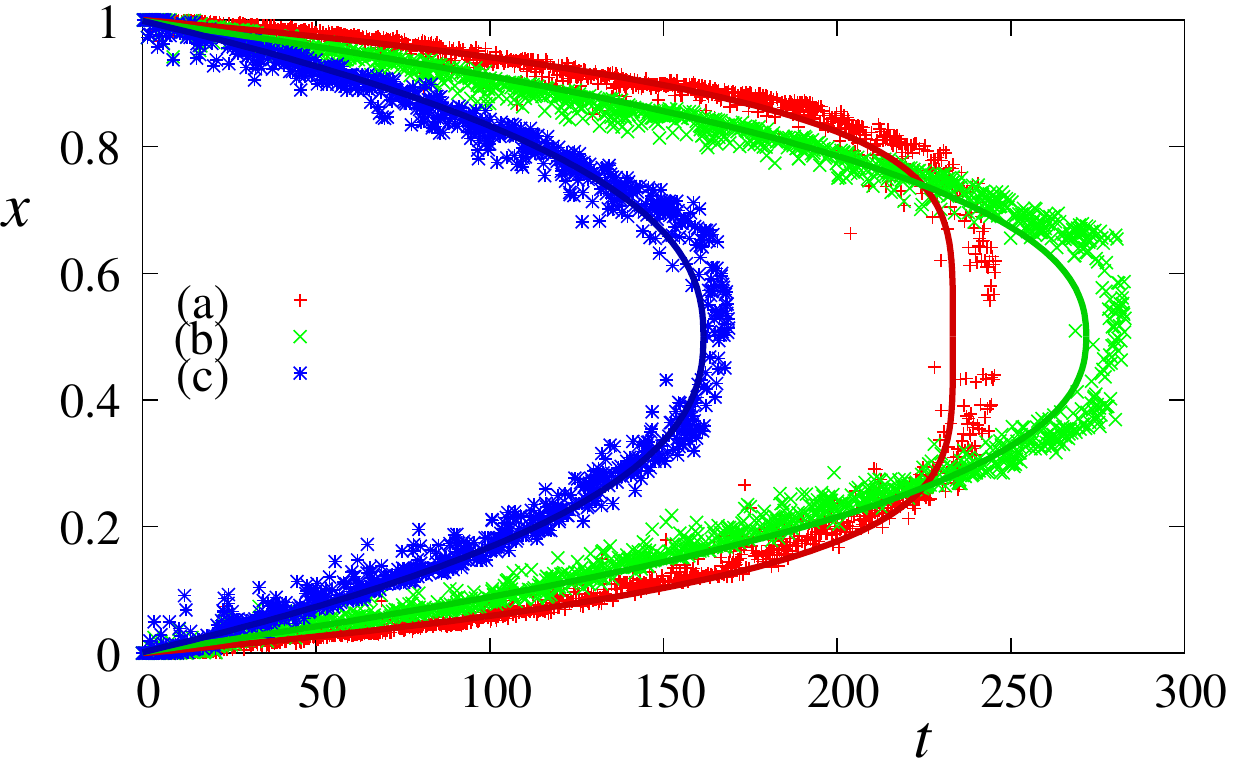}
\caption{\label{Fig:2csopfit} 
The time evolution of the two opinion groups for three sets of parameters: (a) $\ea=0.2$,
$\ma=0.2$, (b) $\ea=0.075$, $\ma=0.5$, (c) $\ea=0.1$, $\ma=0.5$.
The points are simulation results, solid lines are result of
Eq.~(\ref{Eq:2groupanal}).
}
\end{center}
\end{figure}

In order to understand the dynamics of systems with multiple groups,
we start with a two group scenario. The convergence of a
two group system is always fast (see Fig.~\ref{Fig:2csopfit}) and can
be understood by the following reasoning: In the case of two groups the
article can only be found between the two groups. If an agent from
either sides is chosen to edit the article, two scenarios may occur.
Either the article is outside of the tolerance of the agent then the
agent will pull the article towards it's group, in the opposite case
the agent will move towards the article, i.e. towards the center of the
opinion pool. The inter-agent talks keep the opinion groups together
thus if any member moves the whole group follows it though the
distance the group makes will be $n_g$ times smaller ($n_g$ being the
number of agents in the group). Thus the article makes a random walk
between the groups while the groups gradually shift towards each
other.

However, the article does
not make a regular random walk since the step size depends on its
position relative to the groups: the farther it is from the group the
larger the step size it takes in the direction of it. Therefore the random walk of the article is biased towards the center.

The number of steps the article can make from the center at $x_c$
to the group at opinion $x_e$ can be calculated as:
\begin{equation}
n_s(x)\simeq\int_{\ea}^x\frac{1}{\ma y} dy=\frac{1}{\ma}\log(x/\ea),
\end{equation}
where $x=|x_e-x_c|$ is the distance of the two groups (the
number of steps is the integer part of $n_s(x)$ but we approximate it with a continuous variable). 

If the probability of choosing an agent from this group with respect to
the other group is $p_g$ ($p_g=1/2$ for the two group case but the
general scenario will be useful later) then the article is within the
tolerance level of the article if at least $n_s$ number of steps were
made in the direction of the group. The time the article spends in the
vicinity of the extreme group is
\begin{equation}
\tilde t_g\propto p_g^{n_s}.
\end{equation}
The velocity of a group is inversely proportional to the $n_g$ number
of agents it contains, so:
\begin{equation}\label{Eq:veloanal}
\frac{1}{n_g}v_g(x)\propto\frac{\tilde t_g}{n_g}=\frac{1}{n_g}p_g^{\log(x/\ea)/\ma},
\end{equation}
Since we only know $v_g(x)$ we will integrate the inverse of the
latter to get the time as function of the group position
\begin{equation}\label{Eq:2groupanal}
t=\int_0^xp_g^{\log(x'/\ea)/\ma}dx'
\end{equation}
The integral can be calculated for fixed values of $\ma$ and $\ea$.
Figure ~\ref{Fig:2csopfit} compares the analytical result with the simulation
data. The result of Eq.~(\ref{Eq:2groupanal}) fits numerical results
well and gets bad only at the very end where the number of steps the
article can make is small and neglecting the integer part makes an
important error.

In the following we continue with the analysis of the cases with 3 and 4
groups where we can test the effect of the extremists $RoE$ and of the
probability of the communication action $r$.

\begin{figure}
\includegraphics[width=0.95\linewidth]{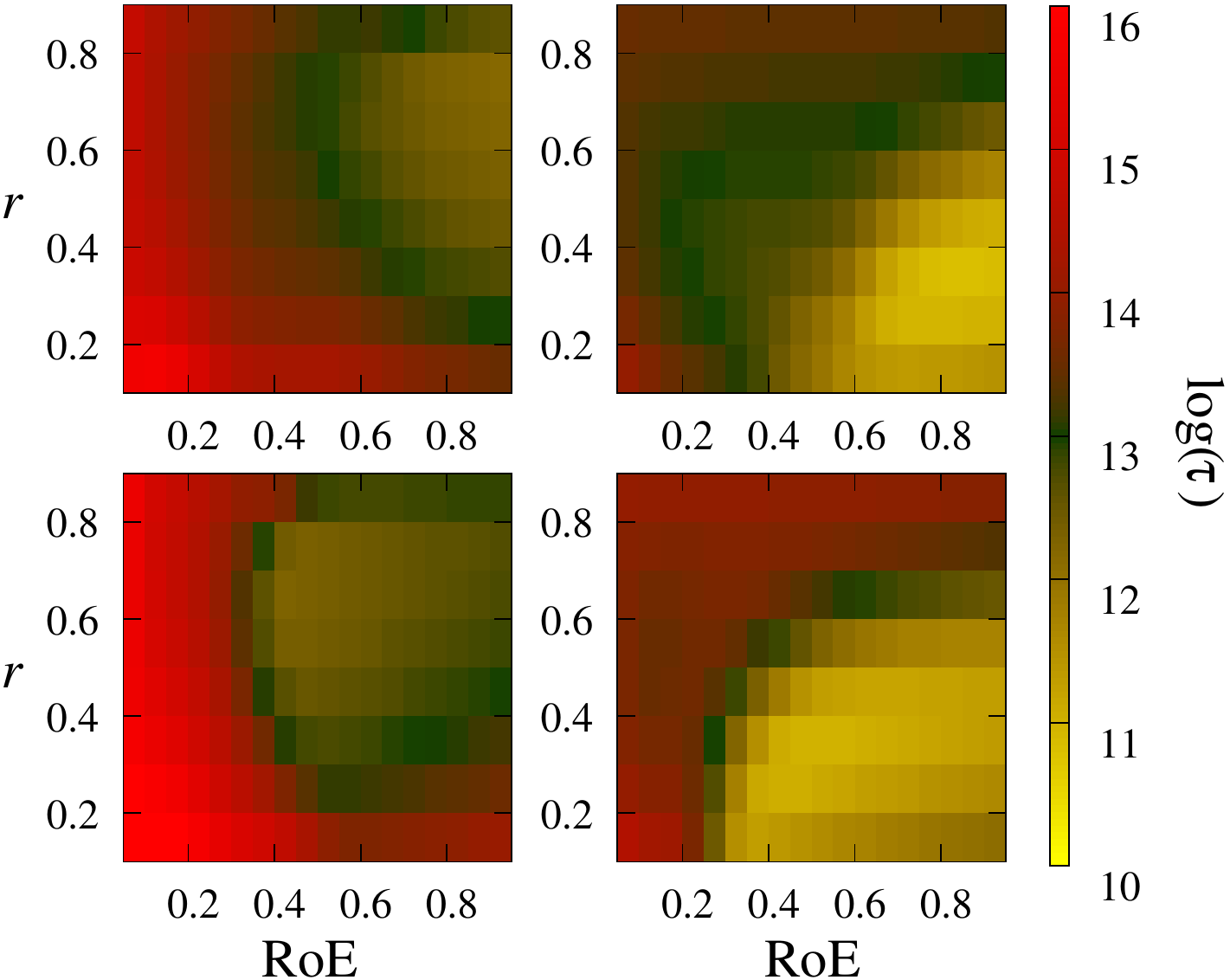}

\includegraphics[width=0.47\linewidth]{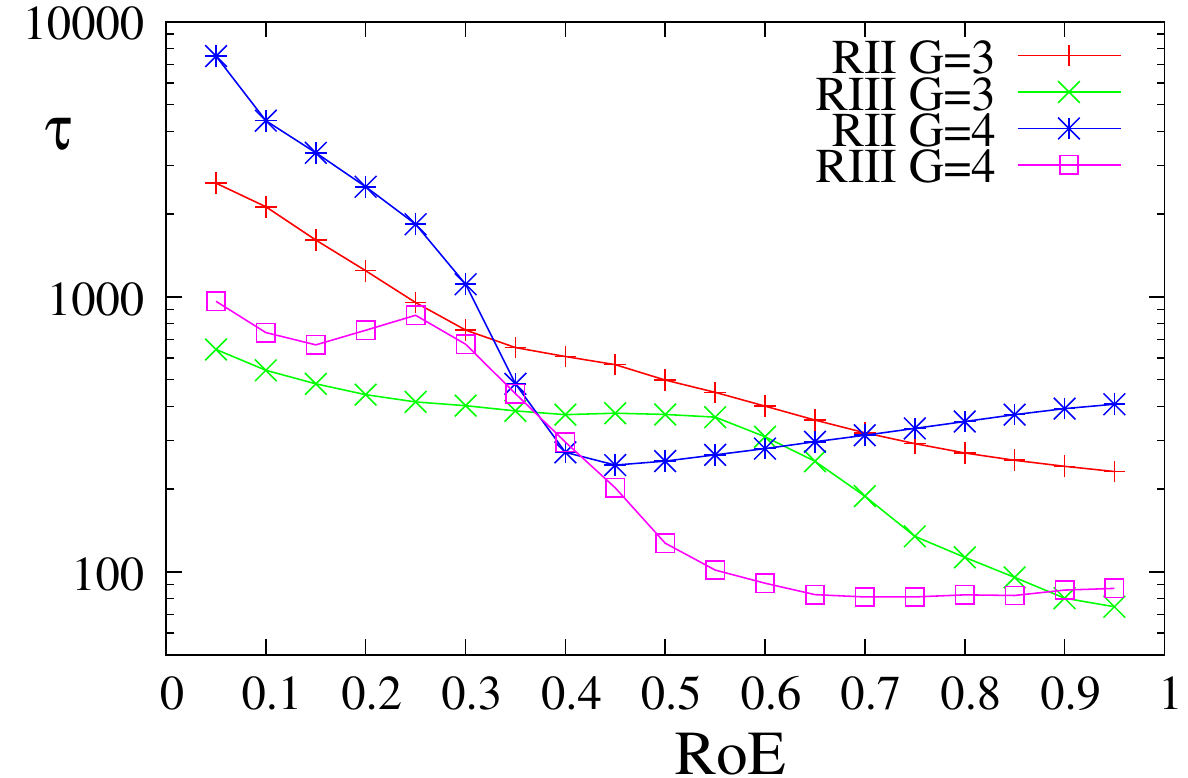}
\includegraphics[width=0.47\linewidth]{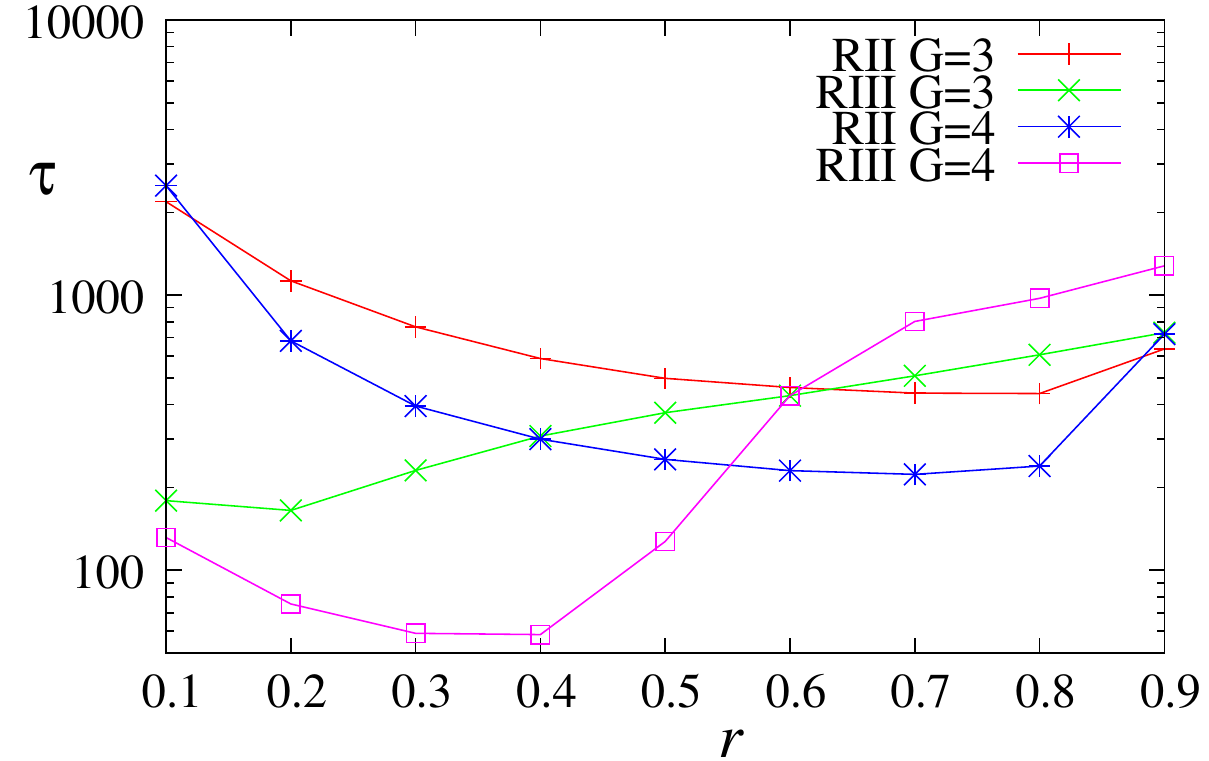}
\caption{\label{Fig:TrRoE} Colormaps of the logarithm of the relaxation time $\tau$ in function of $RoE$ and $r$. Top row 3, bottom row 4 groups, left regime II, and right regime III.
The lower two graphs show cuts at $r=0.5$ and $RoE=0.5$ respectively. Let us note the log-linear scale and the relaxation time sometimes grows an order of magnitude within small changes of $RoE$.}
\end{figure}

\begin{figure}
\includegraphics[width=0.95\linewidth]{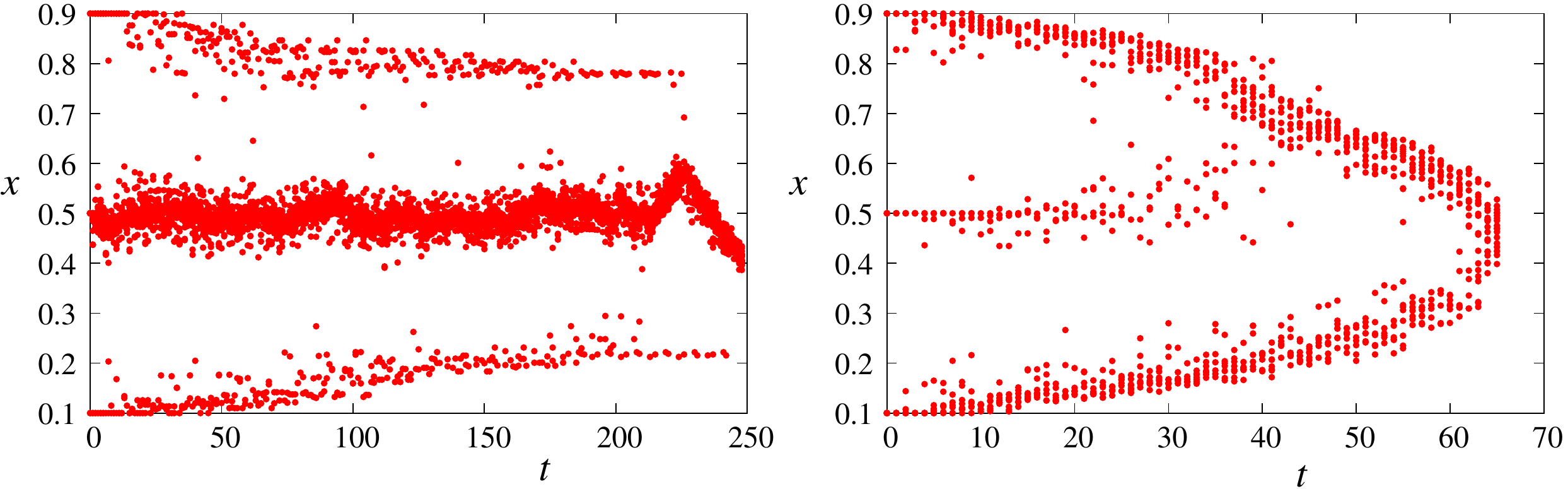}
\caption{\label{Fig:3csopRoE_ex} 
Example time evolution for three initial groups.
$N=1000$, $r=0.5$, $\ea=0.15$, $\ma=0.7$, left: $RoE=0.5$, right: $RoE=0.9$
}
\end{figure}

Figure~\ref{Fig:TrRoE} summarizes the relaxation time as a function of
the two parameters: $RoE$ and $r$. The results are surprising: In case
of three groups the convergence is always faster if the ratio of the
extremists is higher. This unintuitive result can be understood by looking
at specific examples in Fig.~\ref{Fig:3csopRoE_ex}. The high ratio
extremist case behaves essentially as a two group system while on the
other hand if the middle group is strong we obtain the converging
extremes scenario \cite{torok2013} with much larger relaxation time. The transition
is smooth and the minimal relaxation time is observed when there is no mainstream group. This
may seem counter intuitive but in our model the big mainstream group
acts as conflict stabilizer. This issue will be taken up again 
in the Discussion.

\begin{figure}
\includegraphics[width=0.95\linewidth]{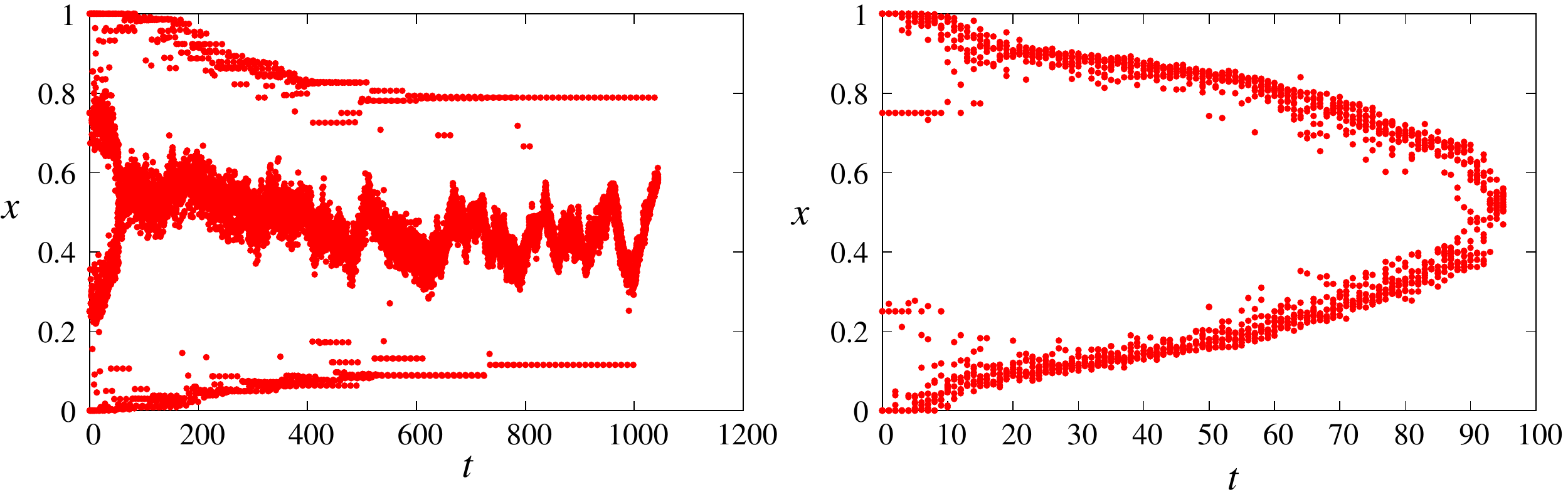}
\caption{\label{Fig:4csopRoE_ex} 
Example time evolution for four initial groups.
$N=1000$, $r=0.5$, $\ea=0.15$, $\ma=0.7$, left: $RoE=0.2$, right: $RoE=0.5$.
}
\end{figure}

Different behavior can be observed in the case of four opinion groups
where, as shown in Fig.~\ref{Fig:TrRoE}, there is an optimum for the relaxation time in function of $RoE$ which
is just after a sharp transition at around $RoE=0.35$. The examples in
Fig.~\ref{Fig:4csopRoE_ex} illustrate the reason behind this
transition. If the middle groups are more numerous than the extreme groups, they merge
fast as a two group system and the system is converted to a big middle
group and two small extremist groups scenario, which is very stable
with long relaxation time. On the other hand, if there are more
extremists, they first merge with the middle group of their side and
thus reducing the four group system to a two group system which converges
fast. The difference in the relaxation time between the two cases is of
orders of magnitude as indicated by the logarithmic scale in
Fig.~\ref{Fig:TrRoE}. If the mainstream groups merge with the
respective extremist then increasing the size of the extremists will
prolonge the debate (see Fig.~\ref{Fig:TrRoE}) because the merged groups
will be farther from the middle opinion.

We identify $RoE_c$ the transition point between the above two scenarios when all four groups merge together simultaneously. We will consider the movement of a mainstreamgroup as the result of independent two group scenarios. Unfortunately, the integral in Eq.~(\ref{Eq:2groupanal}) is impossible to evaluate for the general case, therefore we will evaluate only the initial speed of the group. Since according to Eq.~(\ref{Eq:veloanal}) the smaller groups accelerate faster the transition point should be between two limiting cases: (i) the middle groups start with 0 velocity, and (ii) the middle groups start with third of the extremists group velocities. These two conditions give us two implicite equations which can be solved numerically. The two equations are:
\begin{align}
0=&(1-RoE)v_{1/2}(0.5) -RoE ~ v_{RoE}(0.25) +RoE~v_{RoE}(0.75)\\
\frac{1}{2} v_{1/2}(0.5)=& 2\left[
(1-RoE)v_{1/2}(0.5) -RoE ~ v_{RoE}(0.25)
\right]
\end{align}
The numerical solutions of the above equation give us the range $0.37<RoE_c<0.43$,
which corresponds well to the transition point from the relaxation
times (see Fig.~\ref{Fig:TrRoE}).

Thus, if the mainstream views are divided between two options, as in
many cases normally, then there is an optimal ratio of the extremists.
Of course no extremists would be the best case but as soon as there
are few of them they stabilize the conflict. There comes another
sharp transition when they get numerous enough that the extremist
groups merge with the respective mainstream leading, thus, to a two group
system. The key difference here is that the extremists take an active part
in the debate.

The edit talk ratio $r$ also has an effect on the relaxation time. In regime II for both cases of 3 and 4 groups, there is a medium ($\sim 0.6$) optimum talk/edit ratio leading to the shortest relaxation time. On the contrary, in regime III, low values of $r$ are more favourable for short relaxation time which means a lot of editing and little discussion. In summary, in case of oscillations we need agent-agent discussion for fast convergence but in the converging extremes the major element of the convergence process is the relatively volatile article and extremists adapting a more mainstream opinion due to a semi extremist article. For this we only need editing.

\begin{figure}
\begin{center}
\includegraphics[width=0.85\linewidth]{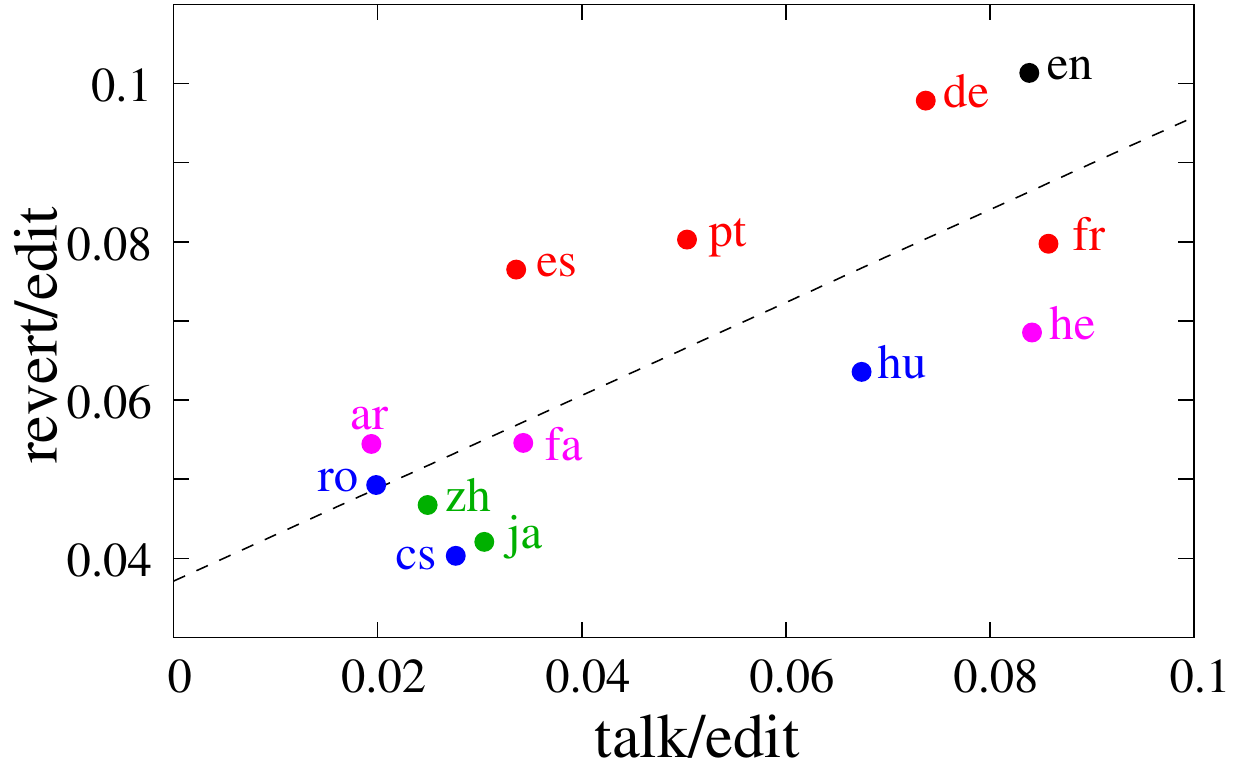}
\caption{\label{Fig:wiki_rm} 
The revert/edit ratio (corresponding to $\tau$) vs. the ratio of edits
to talk/article pages (corresponding to $r$) for 13 different
Wikipedia language editions (en: English, West European (red): de:
German, fr: French, es: Spanish, pt: Portugal, Eastern European
(blue): cs: Czech, hu: Hungarian, ro: Romanian, Middle-East (pink):
ar: Arabic, fa: Persian, he: Hebrew, Far-East (green): zh: Chinese,
ja: Japanese. 
The aim of the dashed line is to drive the eye.
}
\end{center}
\end{figure}

To be able to compare these results with the empirical data, we calculated two ratios: (i) the ratio of edits to talk and to Wikipedia pages and (ii) the ratio of reverts on the articles to Wikipedia page edits for each language edition. One should keep in mind that this ratio is not exactly the same as in the model. In reality, edits to the articles are finer than edits to talk pages which generally add a larger part of the text to the discussion. Hence the talk/edit ratio in the model overestimates the same quantity measures in Wikipedia data. However, we expect to see the same trends. As a proxy to the consensus reaching time, we measured the ratio of reverts to all the edits in each language editions. Fig.~\ref{Fig:wiki_rm} shows the relationship between these two ratios. We observe that in the language editions for which the talk/edit ratio is higher, there are generally more reverts. Similar relationship has also been reported for the article label \cite{yasseri2012dynamics}.
This behaviour is more similar to the case of 3 opinion groups in regime III.

\subsection{Banning of extremists}

In Wikipedia, many different tools are used to control and eventually 
settle conflicts, for
example freezing controversial articles, or banning users who are not obeying community conventions temporarily or permanently. 
The aim of these measures are mainly to calm down
editors, an aspect which is not included in our model. It has been shown that the
editors who are banned more often, have a higher focus on smaller number of articles 
and they contribute directly to the editorial wars \cite{yasseri2013}. 

In our model, we have shown that oscillations have a key role in reaching consensus. Here we investigate how the introduction of banning can effect the consensus reaching process. Our expectation is that by banning agents, the fluctuations (e.g. number of active agents in extremist groups) speed up the oscillations and the relaxation time will be reduced.

We implement banning in our simulations as follows: If an agent in an edit process
changes the article, it may get banned with a probability proportional to
the square of the distance between the opinion of the
agent and of the article. This reflects the fact that agents who find
an article completely opposing their views are more likely to take
action resulting in a banning. The probability of banning can be
formulated as
\begin{equation}
p=\begin{cases}
(A-x_i)^2& \mathrm{if}~ |A-x_i|>\ea\cr
0& \mathrm{otherwise}
\end{cases}
\end{equation}

Banned agents, when selected for editing action will do nothing but get
back their normal status and later will be able to edit the article. Agents are
thus banned for one editing action while they still participate in the
talking.
In the simulation, the following agent opinion distribution
was used ($RoE=0.5$):
\begin{itemize}
\item $N/4$ extremist agents with opinion $[0.05,0.15]$
\item $N/4$ extremist agents with opinion $[0.85,0.95]$
\item $N/2$ mainstream agents with opinion $[0.45,0.55]$
\end{itemize}

We can assume that as soon as the middle group and the article get in the vicinity of an
extremist group, the members of the opposite extremist group get banned
more easily, enhancing thus the oscillatory process of the article. To test this, we
measure the relaxation time ($\tau_b$) and compare it to the case without
banning ($\tau$).

\begin{figure}[ht!]
\includegraphics[width=0.45\linewidth]{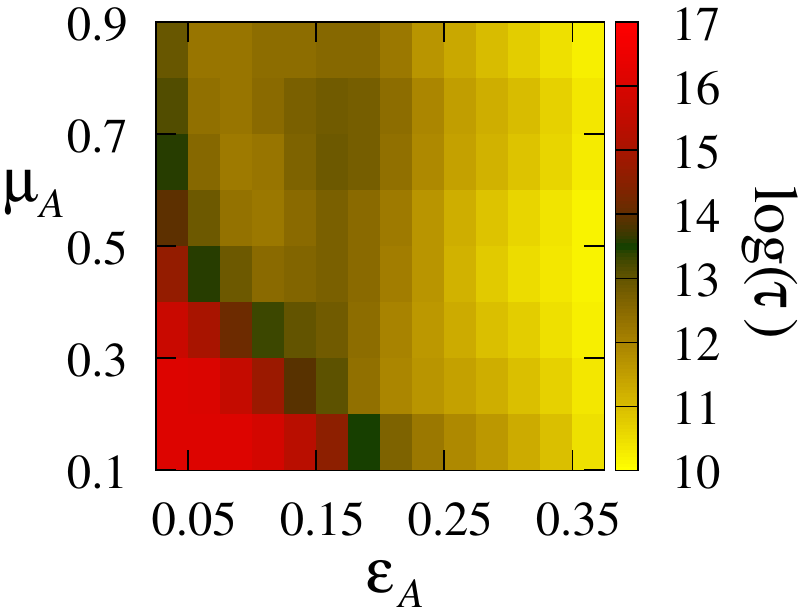}\hspace{0.05\linewidth}
\includegraphics[width=0.47\linewidth]{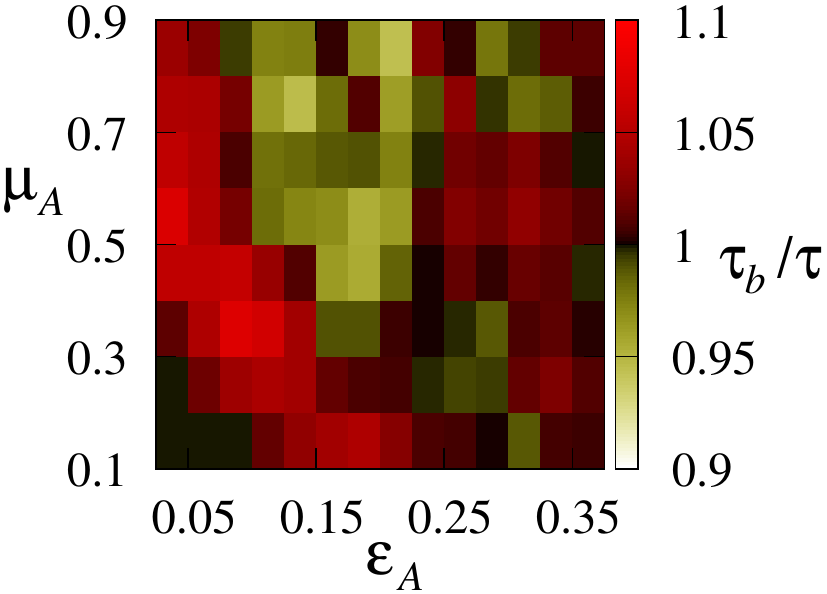}
\caption{\label{Fig:trelax} 
(a) The logarithm of the relaxation time of the original model with
three opinion groups and $\mathrm{RoE}=0.5$, (b) The ratio of the
relaxation time with and without banning.
}
\end{figure}

Figure~\ref{Fig:trelax} shows the logarithm of the relaxation time of
the model without banning, and the ratio of the relaxation time with/
without banning (normal scale). In the presence of banning, in regime III, patches of relaxation time decrease can be
observed, while a more pronounced $\sim10$\% increase is observed in regimes I and
II. Moreover the increase of the relaxation time is larger where the
relaxation time of the original model was already large. Therefore, it
seems that this type of banning may only help to make consensus reaching faster
when it was already fast without. A possible explanation why the relaxation time
increases could be, that when agents are banned, the debate is just
delayed, because with banning it is less likely that a chosen agent is
able to edit. So by banning agents, all we do is slow down the editing
process. It seems, that the effect of positive feedback is too small,
and the fluctuations are not large enough to compensate the loss of actions.

We found that banning users hiders consensus building in most cases.
As already mentioned, in reality in some cases banning may still help
the process of consensus building but for other reasons not included
in the model, for example: banned editors leave the pool, or with
cooler heads after the banning period editors can be more constructive. Thus further improvements on the model could be the implementation of these effects, for example: changing $\ea$ after banning.
But the main message is similar to the previous section:
one needs active interaction of all participants to achieve a
consensus.

\begin{figure}[ht!]
\includegraphics[width=0.95\linewidth]{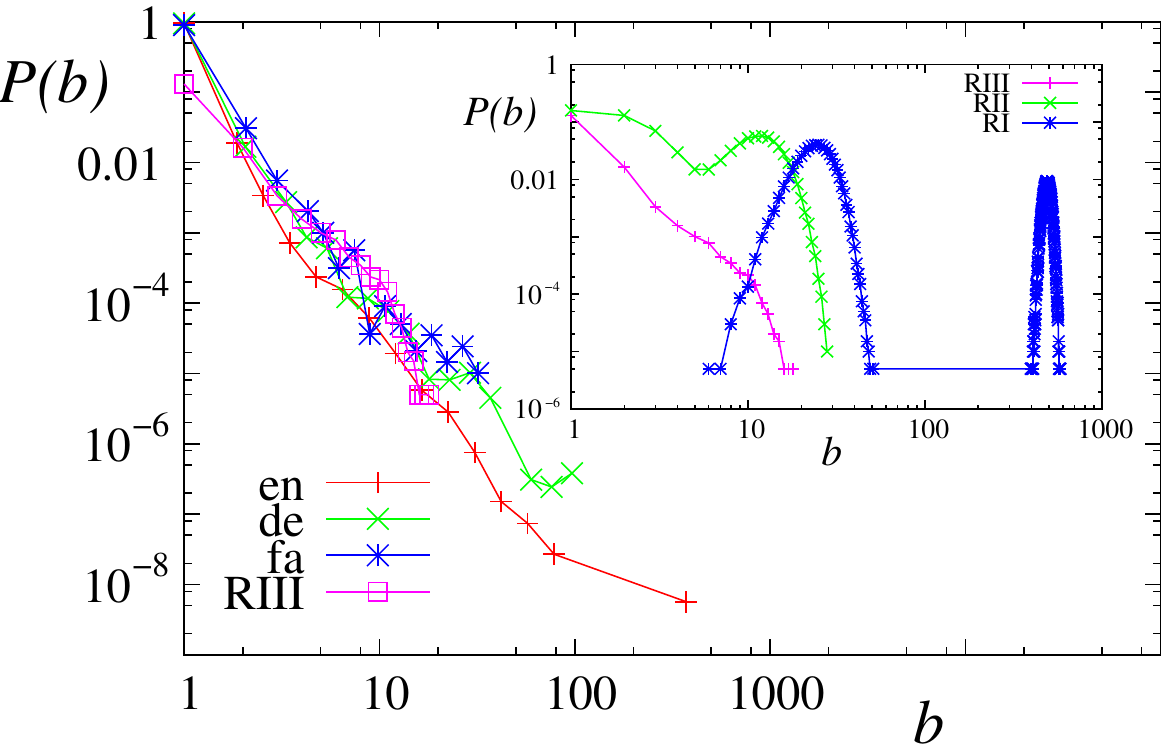}
\caption{\label{Fig:kizaraseloszlas} 
The distribution of the number of times users were banned in different
regimes and in Wikipedia. Notation: en: English, de: German, fa: Persian, RI, RII, RIII stands for Ragime I , II, III respectively. On the main plot only regime III is shown the other regimes for the model are shown in the inset.
}
\end{figure}

It is possible to verify our model by real Wikipedia data as there are
two quantities which can be measured both in the model and on
Wikipedia. 
First we show the distribution of the number of times users
were banned $b$. The results are shown in Fig.~\ref{Fig:kizaraseloszlas}.
The results for regime III look the same as the empirical data
except for users banned once. This small deviation could be explained
by the  vandal editors on the Wikipedia, who  deliberately delete
articles or replace entire pages with nonsense. These editors are only
banned once permanently, a feature (deliberately) missing from our model.

\begin{figure}[ht!]
\includegraphics[width=0.95\linewidth]{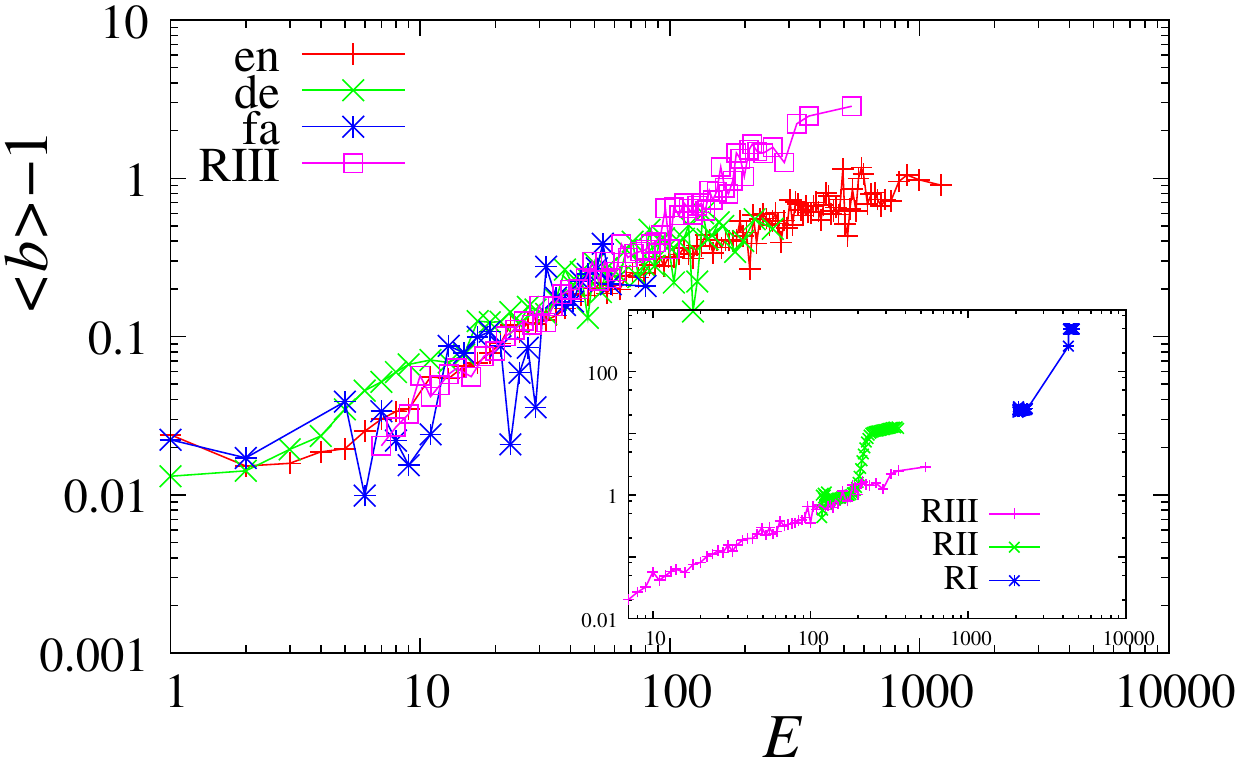}
\caption{\label{Fig:kizarasedit} 
Number of edits versus number of banning (minus 1) in Wikipedia and in
the model. Notation: en: English, de: German, fa: Persian. On the main plot only regime III is shown the other regimes of the model are only shown in the inset.
}
\end{figure}

Now we turn to the relationship between the number of
edits versus the number of bans. The simulation results are compared with actual Wikipedia 
data in
Fig.~\ref{Fig:kizarasedit}. The positive correlation
between the number of edits and the number of times each user is banned reflects the old proverb that "it's only
those who do nothing that make no mistakes". The correlation is
very similar to the empirical observations. The only deviation is for large
number of edits where the model overestimates the number of banns. This
may be the result of the ability that human editors learn how to avoid mistakes
which results in banning, which agents do not do in our model.

Obviously, regime III is the most similar to the Wikipedia data, which
is important since relaxation time is the smallest in this regime
which allows editors to reach a consensus within a reasonable time limit. This could be one reason
behind the sustainability of Wikipedia.

\section{Conclusion}

In this paper we investigated the role of extremists in a value
production environment like Wikipedia using both modeling and data
analysis. Two questions were studied: the influence of the ratio of
the extremists on the characteristic time to reach consensus and the
effect of temporal ban.

We found that in order to achieve fast consensus in our model all
participants need a constructive access to the common product in order
to converge to the consensus. The worst scenario is when there is a
strong mainstream group which punishes all moves of the extremist
groups (a very general scenario in everyday life). We have even found
that there is a phase transition like abrupt change in the relaxation
time in the four opinion group system as function of the ratio of the
extremists and unexpectedly the high relaxation time regime is for low
ratio of extremists.

We have also found out that in general there is an optimal ratio of
talk/edit and it is never zero but especially in Regime III is very
small. This is again a counter intuitive result meaning that more
discussion does not help consensus, very often it just freezes the
front lines. We have shown that in real Wikipedia the articles with
higher talk/edit ratio have relatively more conflict.

Translating it to real life would mean that if debates are mediated by
people with average (politically correct) views then the debate will
not lead to consensus as the extremist groups will remain frustrated
forever. If the extremists have a chance to see their opinion
reflected in the medium for at least a small amount of time they are
more inclined to change their views which is necessary for the
consensus. Thus the active participation of the extremists is needed
for a consensus. Furthermore too much discussion just strengthens the
position of people in their opinion group and does not allow them
to leave it. We believe these observations are well reflected in other
fields of opinion difference e.g. politics.

We also included banning, a general procedure of Wikipedia, and we
found it counterproductive in most cases as it only delays the
consensus building. The probability distribution of the number a user
was banned in Regime III matches very well with the Wikipedia data
suggesting that the converging extremes is the most general
convergence method on Wikipedia. It means that the point of view of
the article is volatile and extremists become staisfied with a
temporary biased article while they also alter their views on the
subject while at the end they accept a more mainstream version of the
article.

\section*{Acknowledgments}

TY has received funding from the European Union's Horizon2020 research and 
innovation program under grant agreement No 645043; "HUMANE: a typology, method and roadmap for HUman-MAchine NEtworks".

\bibliographystyle{plos2015}
\bibliography{banningpaper} 

%
%
%

\end{document}